\documentclass{aa}
\usepackage{graphics}

\begin{document}

\title{The Multicolor Panoramic Photometer-Polarimeter with high time resolution based
on the PSD }

\author{V.Plokhotnichenko \and  G.Beskin \and V.Debur \and A.Panferov \and I.Panferova }

\institute{Special Astrophysical Observatory, Nizhnij Arkhyz, 369167, Russia, Isaac Newton
Institute of Chile, SAO Branch, Russia.}

\date{}

\abstract{
Multicolor Panoramic Photometer-Polarimeter (MPPP) with a time resolution of
1 microsecond has been built based on a PSD and used at the 6-meter telescope
in SAO (Russia). The device allows registration of the photon fluxes in four
photometric bands simultaneously and finding values of 3 Stokes parameters.
MPPP consists of Position-Sensitive Detector (PSD), acquisition MANIA-system,
polarization unit and a set of dichroic filters. MPPP gives a possibility of
detecting photons in 2 pupils with a size of 10 - 15 arc sec centered on the
object and comparison star positions simultaneously. The first half of the object
photon flux passes through the phase rotating plate and polarizer, and the second
one through the polarizer alone. MPPP registers in each of the 4 filters four
images of the object with different orientations of polarization plane and one
image of a comparison star. It allows measuring instantaneous Stokes parameters.
The main astrophysical problems to be solved with MPPP are as follows: investigation
of optical pulsars; study of GRB phenomenon in the optical range; searching
for single black holes; study of fast variability of X-ray binaries. As an illustration
of MPPP use, the results of observations at the 6-meter telescope of Crab pulsar
and soft gamma repeater are presented. 
}

\authorrunning{Beskin et al}
\titlerunning{Multicolor Panoramic Photometer-Polarimeter}
\maketitle

\section{Introduction}

For studying fast brightness variations of faint astrophysical objects it is
necessary to use panoramic detectors of high time resolution. Detectors of such
a type determine both coordinates and arrival time of each photon. The S/N ratio
in this case reaches its maximum value at any seeing and sky background level.
As a rule, the PSD which registers separate photons is created on the basis
of the standard photocathode with low (relative to CCD matrices) quantum output,
a set of microchannel plates and a position sensitive anode (Debur et al., 2002,
this Conference). Some evident shortcomings areas follows: 

\begin{itemize}
\item Low sensitivity (5-20\%); 
\item Narrow dynamical range (limiting flux is 50-500 thousand photocounts/s); 
\item Low spatial resolution. 
\end{itemize}

To minimize their influence when designing photometric detectors, it is desirable
to use the following techniques: 

\begin{itemize}
\item Simultaneous registration of emission in different spectrum regions (with different
orientations of the polarization plane) with one PSD; this is achieved by using
a set of dichroic light dividers;  
\item The registration of the object and comparison star images with their close neighbourhood
(10''- 15'') instead of 1'-2';  
\item Use of antireflecting field reducers varying the scale to 0''.2 - 0''.3 per
element of resolution. The principles mentioned above form a foundation for
construction of panoramic photometer- polarimeter for the 6-m telescope.
\end{itemize}

Even in spite of existence of more sensitive deteoctors like CCD, Postition-Sensitive
Detector still has its own applications. It may be used for study of 
fast variability (with up to several microseconds resolution) of
faint sources (for example it is ibpossible to analyse $19^m$ star variability
using CCD with 1 count per pixel read-out noise). 
Comparison with cryogenic detectors developed now also shows that their are
much more expensive and diffucult to use and provide much worse quantum efficiency.

\section{Peculiarities of the optical scheme}

\begin{figure}
{\centering \resizebox*{1\columnwidth}{!}{\includegraphics{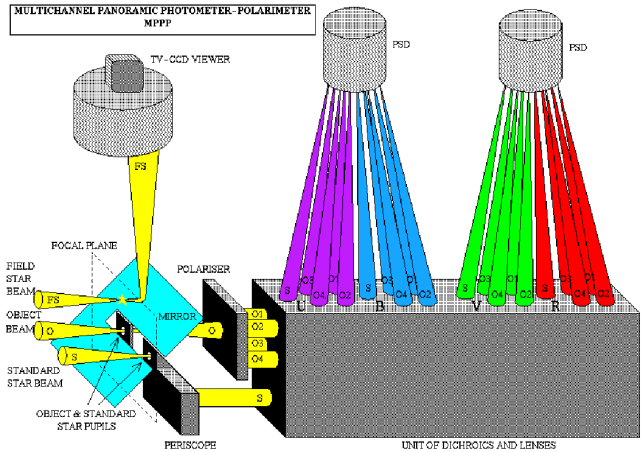}} \par}
\caption{General view of the optical layout.}
\label{fig: optic}
\end{figure}

Panoramic photometer-polarimeter registers only the light fluxes of the object
under study and of the comparison star with their close vicinities, Fig.\ref{fig: optic}.
The light beam from the object is decomposed by the polarization unit into 4
components with different orientations of the polarization plane. The light
beams from the object and comparison star pass through the set of dichroic filters
and are registered with two PSD, one of which is optimized for the blue and
the other for the red regions of the visible spectrum. As a result, each beam
can be registered in four color bands U, B, V, R. For identification of the
object being studied and precise setting it on the data registration channel,
we employ the stellar field viewing based on a TV CCD camera which enables image
acquisition on its matrix and fulfilling a sufficiently deep survey of the selected
sky area. Such a scheme obviates the problem of registration of an excess flux
of quanta and allows fulfilment of a multi-mode analysis of the fast-variable
source emission.

\section{Data registration}

\begin{figure}
{\centering \resizebox*{1\columnwidth}{!}{\includegraphics{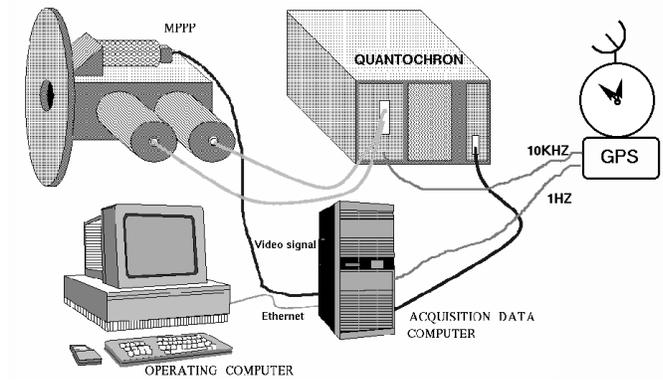}} \par}
\caption{The data acquisition complex.}
\label{fig: comlex}
\end{figure}

The photometer-polarimeter operates as part of the acquisition complex incorporating
also a device for receiving photocount flux codes, a computer for control and
data acquisition, which is located in the local net with the computer of the
astronomer- operator. The functional diagram of the photometrical complex is
shown in Fig. \ref{fig: comlex}. The flux of photocounts from the PSD must
be transferred to the computer data storage in the form it is received by the
detector, as counts of registration time of all the quanta and their coordinates,
the rate of count arriving must be up to 100 000 quants/s. For this purpose,
we use the time-code converter Quantochron {[}1{]}. Primarily it was designed
for registration of the data flows which have a 16-bit coordinate field (265x256
elements). The application of the PSD with a quadrant collector and the use
of 10-bits ADC demanded extension of capacity up to 40 bits, this is why as
a temporal measure we use a multiplexor which disconnects each arriving 40-bits
photocount into three sequential messages: 8, 16 and 16 bits. The 8 bits in
the first message are used for auxiliary information. The data obtained are
stored and transferred by the 100mb Ethernet from the data acquisition computer
to the control computer and then they are written on magnetic disks and then
on CDROM. Storage of all primary data on the measured charges of each photocount
allows a deeper analysis of data flow as to the presence and compensation for
the instrumental effects.

\section{Analysis methods}

\begin{figure}
{\centering \resizebox*{1\columnwidth}{!}{\includegraphics{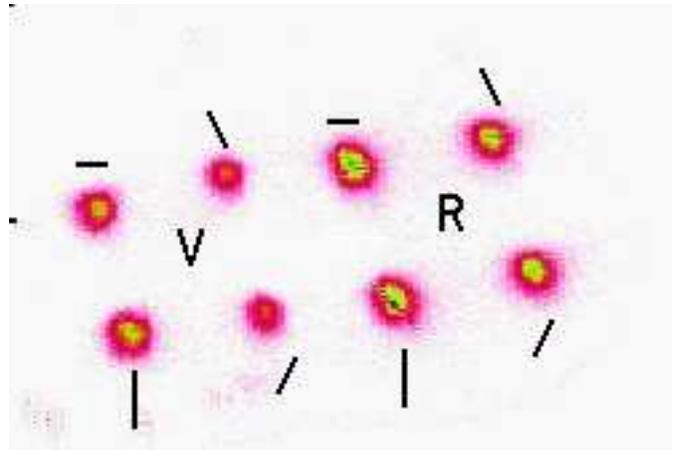}} \par}
\caption{Multimode image of the standard star in the color bands B,
V and 4 orientations of the polarization plane.}
\label{fig: BD28}
\end{figure}

Observational data files are used, first of all, for obtaining images supplied
to the detector cathode by the optical system of the photometer-polarimeter.
As an example, in Fig. \ref{fig: BD28} we show the field of the detector with
the standard star located on it in two-colour bands and four polarizations. 

\begin{figure}
{\centering \resizebox*{1\columnwidth}{!}{\includegraphics{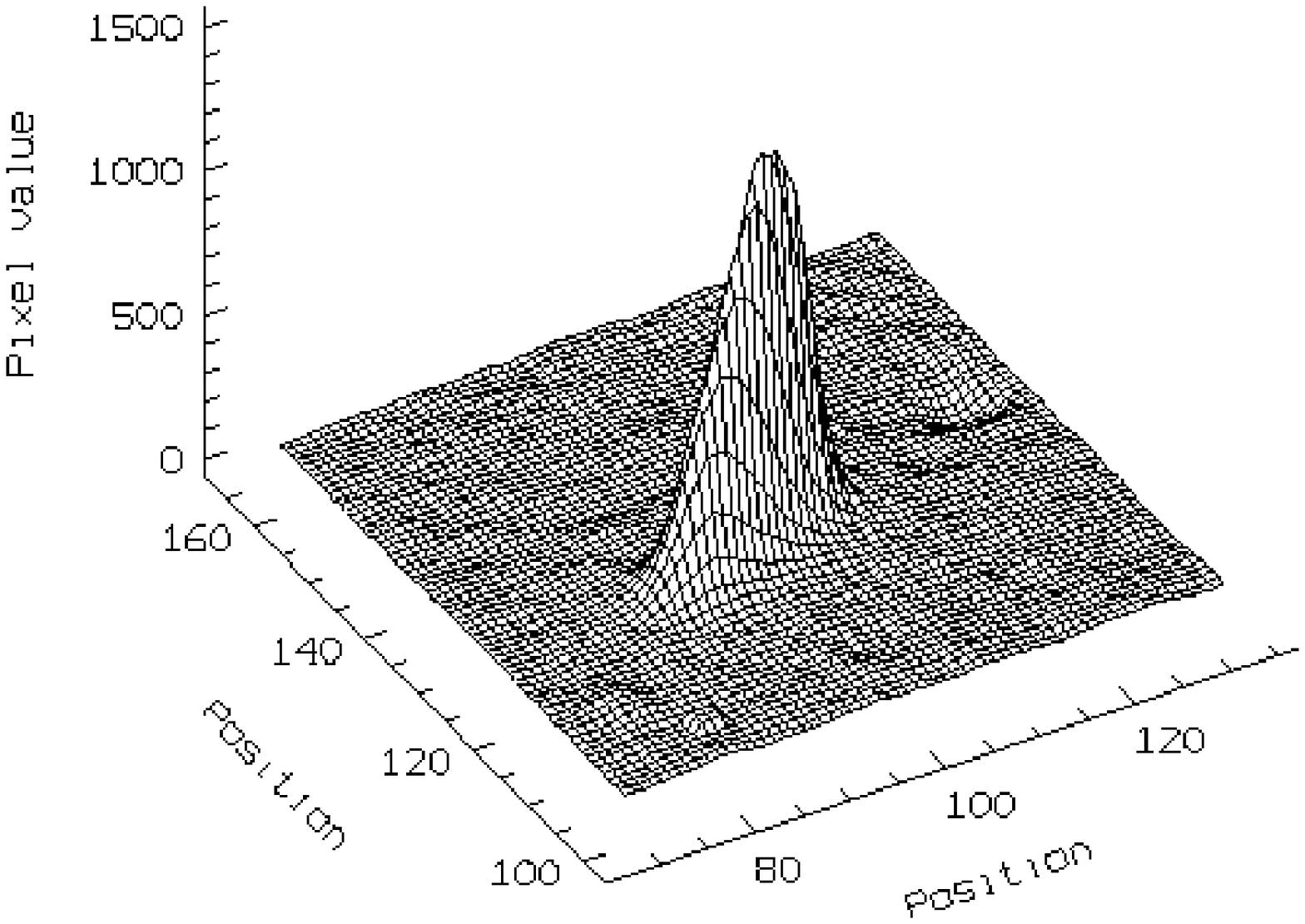}} \par}
{\centering \resizebox*{1\columnwidth}{!}{\includegraphics{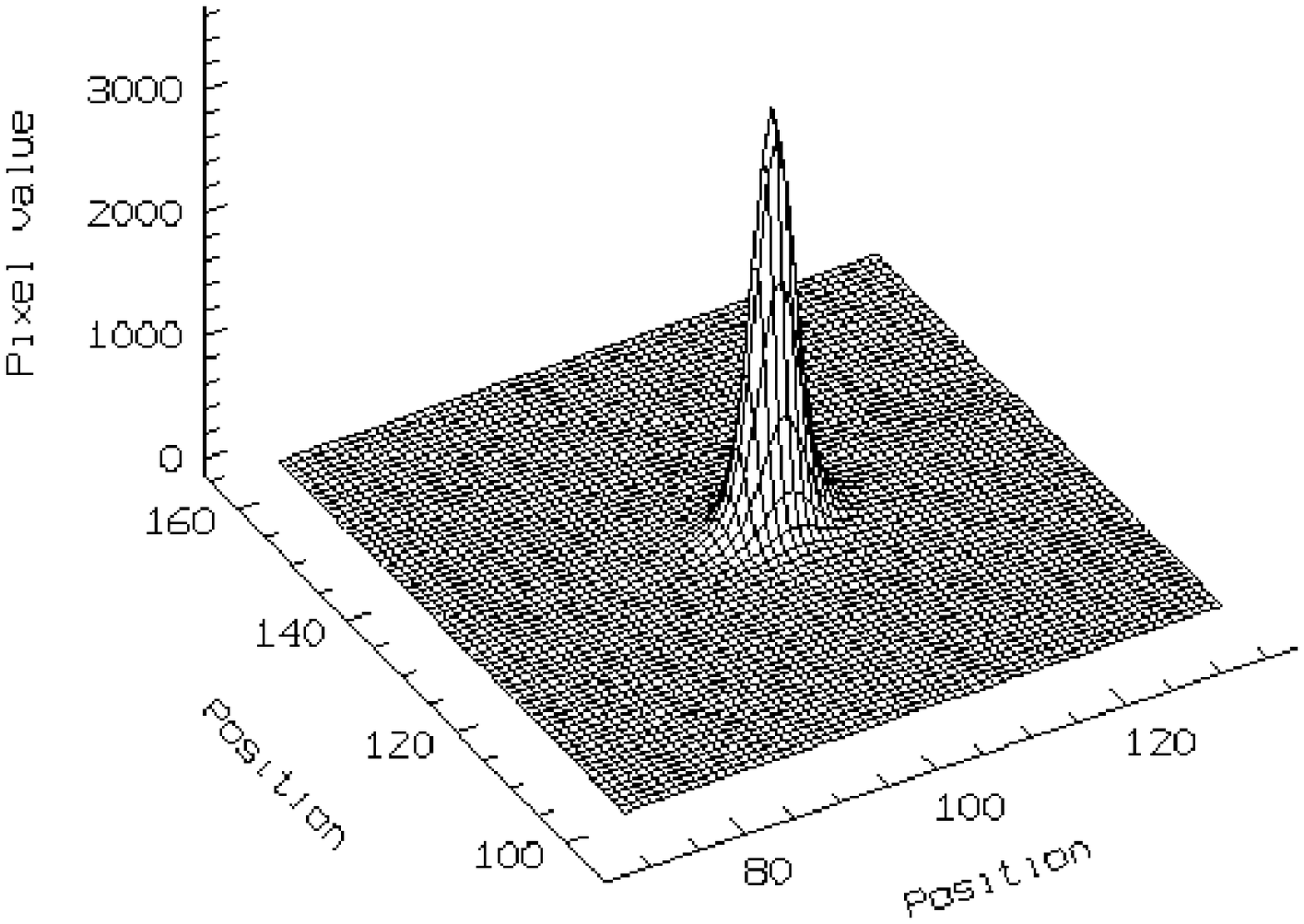}} \par}
\caption{Improvement of stars seeing and signal/noise ratio by the deconvolution
procedure.}
\label{fig: HZ}
\end{figure}

The quantum fluxes from the star in every position which are received with the
PSD are subjected to temporal analysis for studying the processes occurring
on the astrophysical object under study. Prior to the procedure of temporal
analysis we improve the quality of the image produced by the astronomical optics
and improve the S/N ratio by means of determination of coordinate variations
of the integral center of the comparison star image on the detector and adaptive
adjusting of the current location of the area for photocount flux selection,
fig. \ref{fig: HZ}.

\begin{figure}
{\centering \resizebox*{1\columnwidth}{!}{\rotatebox{270}{\includegraphics{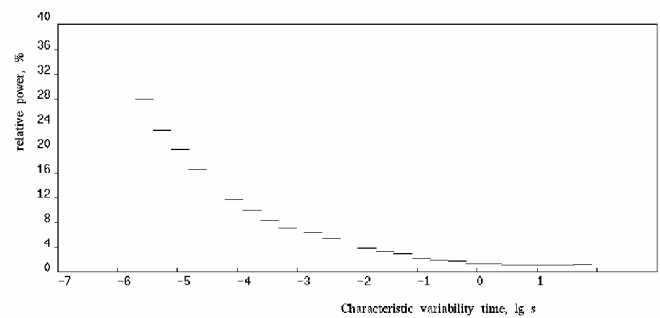}}} \par}
\caption{Upper limits of the variable power component relative to a
background level of soft gamma repeater SGR 1806-20. Restrictions correspond
to stochastical triangle-like flares with a filling factor of 0.1 and a confidential
probability of 99\%. They will be \protect\( 3\protect \) and \protect\( 3\cdot 10^{3}\protect \)
times lower for the filling factors \protect\( 10^{-2}\protect \) and \protect\( 10^{-6}\protect \),
respectively.}
\label{fig: lim}
\end{figure}

\begin{figure}
{\centering \resizebox*{0.6\columnwidth}{!}{\includegraphics{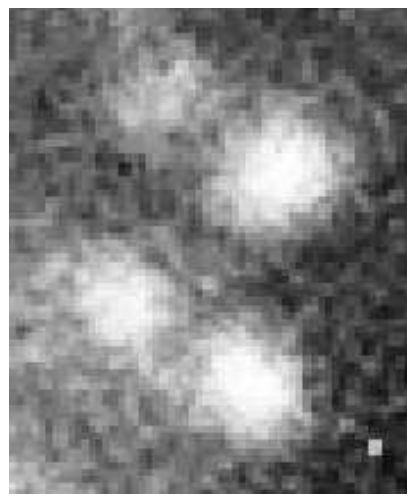}} \par}
\caption{Images of Crab pulsar (left) and standart star (right) in
2 orientations of the polarisation plane.}
\label{fig: crab}
\end{figure}

We use the methods of search for stochastic variability in the range from microseconds
to the exposure duration {[}2{]}. As example in fig. \ref{fig: lim} is shown
result of search for stochastic brightness variability of soft gamma repeater
SGR 1806-20 {[}3{]}. So we search for and analyse periodical variability different
astrophysical objects, for example, optical pulsars. In fig.\ref{fig: crab}
we demonstrate the images of the Crab pulsar and star-neighbour splited by polariser.
Their phase- resolved images and the Crab pulsar folded light curve with a time
resolution of 30 mcs are shown in fig. \ref{fig: lim}. 

\begin{figure}
{\centering \includegraphics{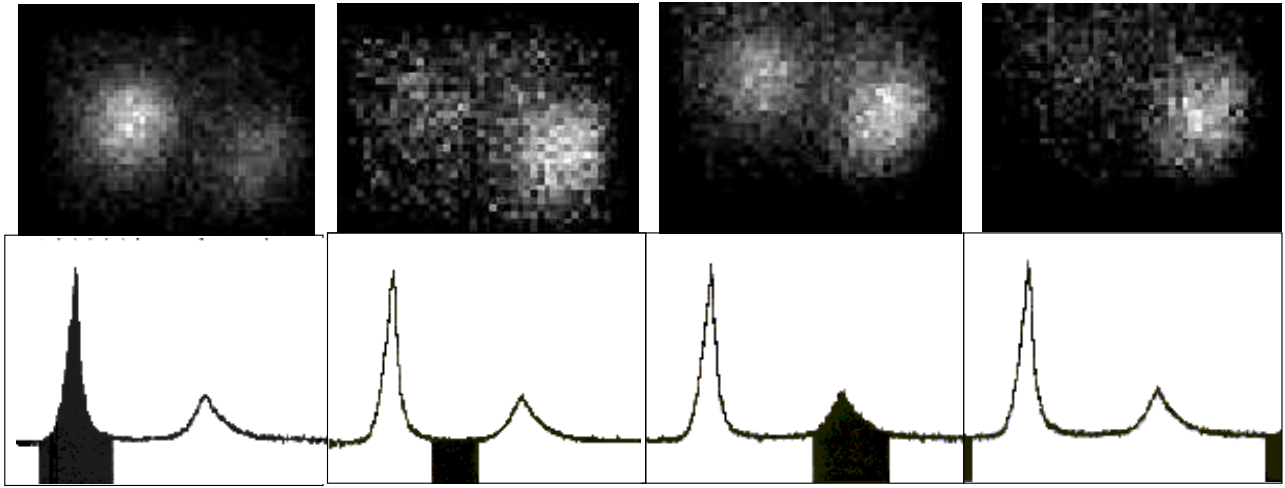} \par}
\caption{Phase-resolved two dimension photometry of the inner Crab nebula in the B band,
taking within radius of 50 pixels, the pulsar being center star. The location
of each phase region is indicated in the light curve obtained from a radius
of 15 pixels from Crab centroid, with the accompanying photometric image}
\end{figure} 

\acknowledgements{
This investigation was supported by the Russian Ministry of Science, Russian
Foundation of Basic Researches (grant 01-02-17857), Federal Program \char`\"{}Astronomy\char`\"{},
INTAS (grant 96-542) and CRDF (grant RP1-2394-MO-02).
}

\end{document}